\documentclass[manuscript]{aastex}

\newcommand{\HST}{{\it HST}}

\newcommand{\kms}{{\>\rm km\>s^{-1}}}

\begin{document}

\title{{\em Hubble Space Telescope\/} Imaging of the Site of the Red Transient
V4332~Sagittarii\altaffilmark{*}}

\author{
Howard E. Bond\altaffilmark{1,2}
}

\altaffiltext{*}
{Based on observations made with the NASA/ESA {\it Hubble Space
Telescope}, obtained by the Space Telescope Science Institute. STScI is operated
by the Association of Universities for Research in Astronomy, Inc., under NASA
contract NAS5-26555.}

\altaffiltext{1}
{Department of Astronomy \& Astrophysics, Pennsylvania State
University, University Park, PA 16802, USA; heb11@psu.edu}

\altaffiltext{2}
{Space Telescope Science Institute, 
3700 San Martin Dr.,
Baltimore, MD 21218, USA}

In recent years several classes of luminous optical transients have been
discovered that are characterized by dust formation and extremely red colors as
their eruptions proceed. Their maximum luminosities lie between those of
classical novae and supernovae (SNe). One subclass of these
events---``intermediate-luminosity red transients'' (ILRTs)---appear to arise
from fairly massive stars, since they are found in spiral arms of nearby
galaxies. Well-known examples include NGC~300 OT2008-1 (Bond et al.\ 2009;
Humphreys et al.\ 2011), and a near-twin of this event, SN~2008S in NGC~6946
(e.g., Szczygie{\l} et al.\ 2012 and references therein). Archival infrared
images of the host galaxies revealed that both of these events arose from
massive, dust-enshrouded progenitors that had been on or near the asymptotic
giant branch just before the explosions.  A possibly related class of dusty
transients is the recently discovered ``SPRITE''s, which reach luminosities
similar to ILRTs, but lack counterparts at optical wavelengths (Kasliwal et al.\
2017). 

Another subset of red optical transients with lower luminosities than
ILRTs---sometimes called ``luminous red novae''---generally arise from older
stellar populations. Examples of this subclass are V838~Mon (which illuminated a
spectacular light echo; e.g., Bond et al.\ 2003; Sparks et al.\ 2008), the M31
Red Variable (M31~RV; Bond 2011 and references therein), and NGC 4490-OT2011
(Smith et al.\ 2016). These events have been attributed to stellar mergers, as
suggested by the multi-peaked outburst light curves of V838~Mon and NGC
4490-OT2011. Strong support for the merger scenario came from the discovery that
the Galactic red transient V1309~Sco was demonstrably a close binary before its
outburst (Tylenda et al.\ 2011). For a recent review of red transients, see 
Kami{\'n}ski et al.\ (2018)

Here I report optical imaging of the red optical transient V4332 Sgr, obtained
in 2014 with the Wide Field Camera~3 (WFC3) on the {\it Hubble Space
Telescope\/} (\HST)\null. V4332~Sgr erupted in 1994 in the direction of the
Galactic bulge, and rapidly evolved from spectral type K to M, and then late~M
(Martini et al.\ 1999). Soker \& Tylenda (2003) proposed a stellar-merger origin
for this outburst. Kami{\'n}ski \& Tylenda (2011, 2013) argued that the central
star, hypothetically the merged single star, is now obscured in a dusty disk,
viewed approximately edge-on. They presented optical imaging and spectroscopic
polarimetry, showing a large degree of linear polarization consistent with a
considerable fraction of the light being starlight scattered off the surrounding
dust. A nominal angular diameter of the dust shell in 2014 can be predicted as
follows: Martini et al.\ reported a (spectroscopic) expansion velocity of about
$v_{\rm exp} \simeq 100\kms$, and Kami{\'n}ski et al.\ (2010) argued that the
distance to V4332~Sgr is at least 1~kpc, and adopted 1.8~kpc. If the ejection
was launched at the outburst near the beginning of 1994, its angular diameter
would be given by 
$$ {\rm diam}
\simeq 0\farcs44 \left({v_{\rm exp}\vphantom{d}}\over{100 \kms\vphantom{p}}\right)
\left({1.8\,\rm kpc\vphantom{d}}\over{d\vphantom{p}}\right)
\left({t-1994\vphantom{d}}\over{2014-1994\vphantom{p}}\right) \, , $$ 
where $d$ is the distance, and $t$ is the year of observation. Thus, it was
expected that the dust shell around V4332~Sgr would be resolvable by \HST\/ at a
time about 2~decades following the outburst.

My \HST\/ observations were obtained on 2014 November~12 (program ID GO-13851),
using the WFC3 UVIS channel (plate scale $0\farcs040\,\rm pixel^{-1}$). I
employed a $512\times512$ pixel ($20\farcs3\times20\farcs3$) subarray with a
three-point dithering pattern, and the $B$ (F438W), $V$ (F606W), and $I$ (F814W)
filters. The source is extremely red, and I used exposure times of $3\times582$,
$3\times24$, and $4\times5$~s, respectively. There is also one set of archival
\HST\/ images available for V4332~Sgr, obtained on 1997 November~3 with the
WFPC2 camera and the narrow-band H$\alpha$ (F656N) filter with an exposure of
$2\times160$~s (GO-7386, PI F.~Ringwald). 

In the top panel of Figure~1 I show false-color representations of the four
\HST\/ images. The 1997 frame shows an essentially stellar image profile.
However, the 2014 images clearly resolve the ejecta in all three filters. They
show an elongated structure, extending from the north-northeast to the
south-southwest. An eye estimate of the dimensions of the nebula in the $B$
bandpass is about $0\farcs53\times0\farcs42$ (including a small correction for
the FWHM of stellar images in the $B$ frames of $0\farcs088$). These dimensions
agree remarkably well with the expectation from the equation given above, and
appear to confirm that we are seeing ejecta from the 1994 eruption. In the
bottom image in Figure~1 I show a color rendition of the 2014 images, with red,
green, and blue encoding the $I$, $V$, and $B$ frames, respectively. This image
shows that the ejecta are bluer than the central source, consistent with
dust-scattered light.

Kami{\'n}ski et al.\ (2018) recently presented submillimeter imaging and
spectroscopy of V4332~Sgr, obtained with the Atacama Large Millimeter\slash
submillimeter Array (ALMA) and the Submillimeter Array (SMA). Their maps of CO
emission (their Fig.~4 in particular) are remarkably similar in morphology,
orientation, and size to the \HST\/ images.  

Kami{\'n}ski et al.\ (2018) argue for a considerably larger distance of about
5--5.5\,kpc, based on an expansion parallax of the CO emission. Unfortunately,
the recent {\it Gaia\/} DR2 (Gaia Collaboration et al.\ 2018) does not provide a
useful parallax for V4332~Sgr ($\pi=0.0017\pm0.2798$~mas), presumably because of
its non-stellar morphology. The DR2 gives a proper motion of
$(\mu_\alpha,\mu_\delta)=(-2.805\pm0.611,-5.398\pm0.641)\,\rm mas\,yr^{-1}$. It
is of possible interest that a 12th-mag star, lying just $10''$ south of
V4332~Sgr, has the same proper motion within the uncertainties:
$(\mu_\alpha,\mu_\delta)=(-2.824\pm0.072,-5.941\pm0.067)\,\rm mas\,yr^{-1}$. The
DR2 radial velocity of this star is $-81.53\pm0.89\,\kms$, which is similar to
that of the molecular emission reported by Kami{\'n}ski et al.
This star does have a precise DR2 parallax, $\pi=0.6121\pm0.0430$~mas
($d\simeq1.63\pm0.12$~kpc), which is very close to the value assumed in the
equation above. 

V4332~Sgr reached a maximum brightness of $V\simeq8.5$ (Martini et al.\ 1999). 
The reddening in the direction of V4332~Sgr, from the online
tool\footnote{{\tt\url{http://argonaut.skymaps.info/query}}} of
Green et al.\ (2018), is about $E(B-V)\simeq0.32$ for distances up to
$\sim$4~kpc, rising to 0.41 at larger distances. Assuming a parallax of
0.6121~mas, and this amount of reddening, the absolute visual magnitude at
maximum was only $M_V\simeq-3.6$. At 5~kpc, the absolute magnitude at maximum
would be $M_V\simeq-6.3$, which is still considerably fainter than the values
reached by V838~Mon or M31~RV. 

\acknowledgments

Support for Program number GO-13851 was provided by NASA through a grant from
the Space Telescope Science Institute, which is operated by the Association of
Universities for Research in Astronomy, Incorporated, under NASA contract
NAS5-26555.

This work has made use of data from the European Space Agency (ESA) mission {\it
Gaia\/} (\url{{\tt https://www.cosmos.esa.int/gaia}}), processed by the {\it
Gaia\/} Data Processing and Analysis Consortium (DPAC, \url{{\tt
https://www.cosmos.esa.int/web/gaia/dpac/consortium}}). Funding for the DPAC has
been provided by national institutions, in particular the institutions
participating in the {\it Gaia\/} Multilateral Agreement.

{\it Facilities:} 
\facility{HST (WFC3)}

\begin{figure}
\begin{center}
\includegraphics[height=4.15in]{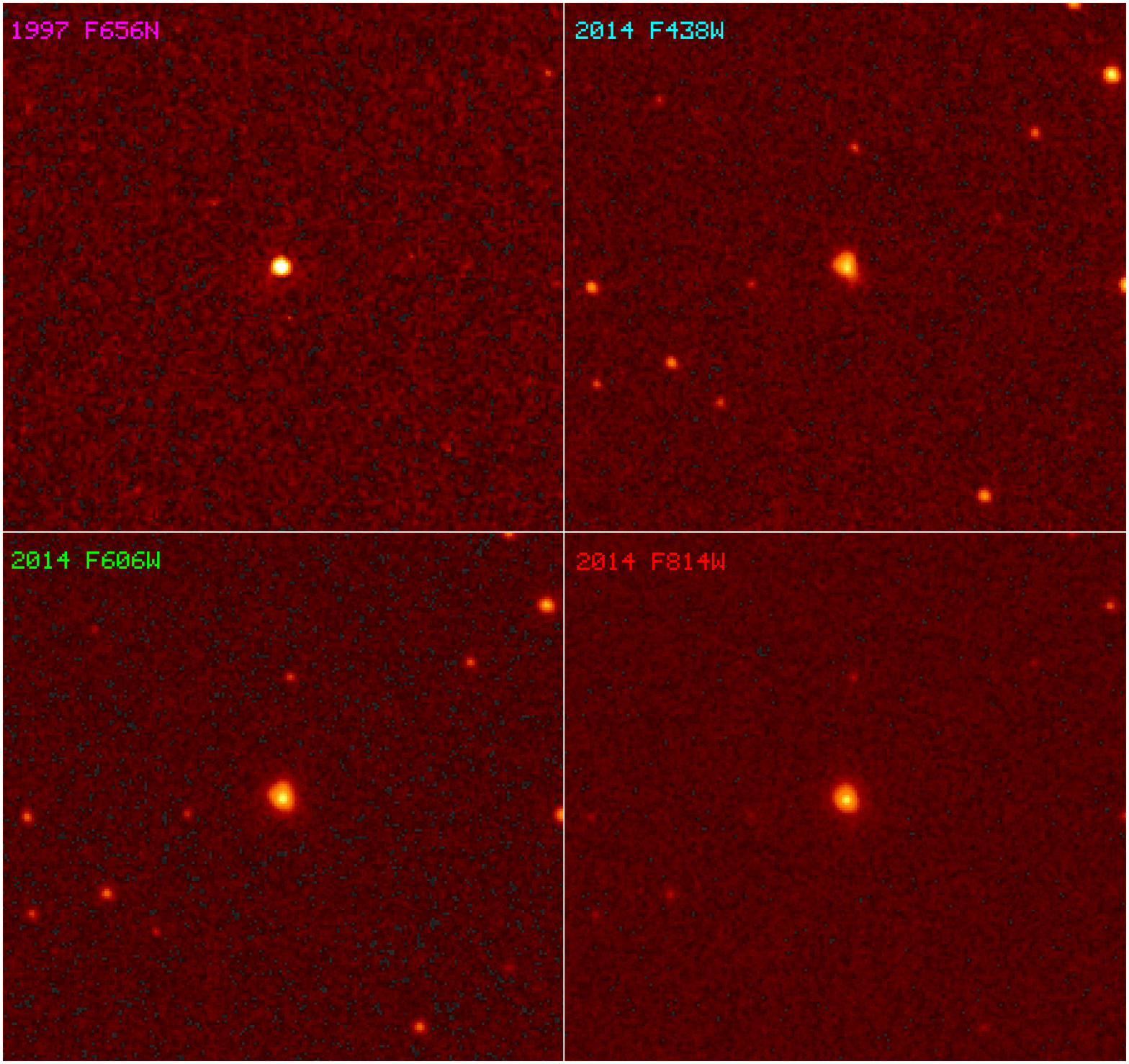}
\vskip0.1in
\includegraphics[height=2.35in]{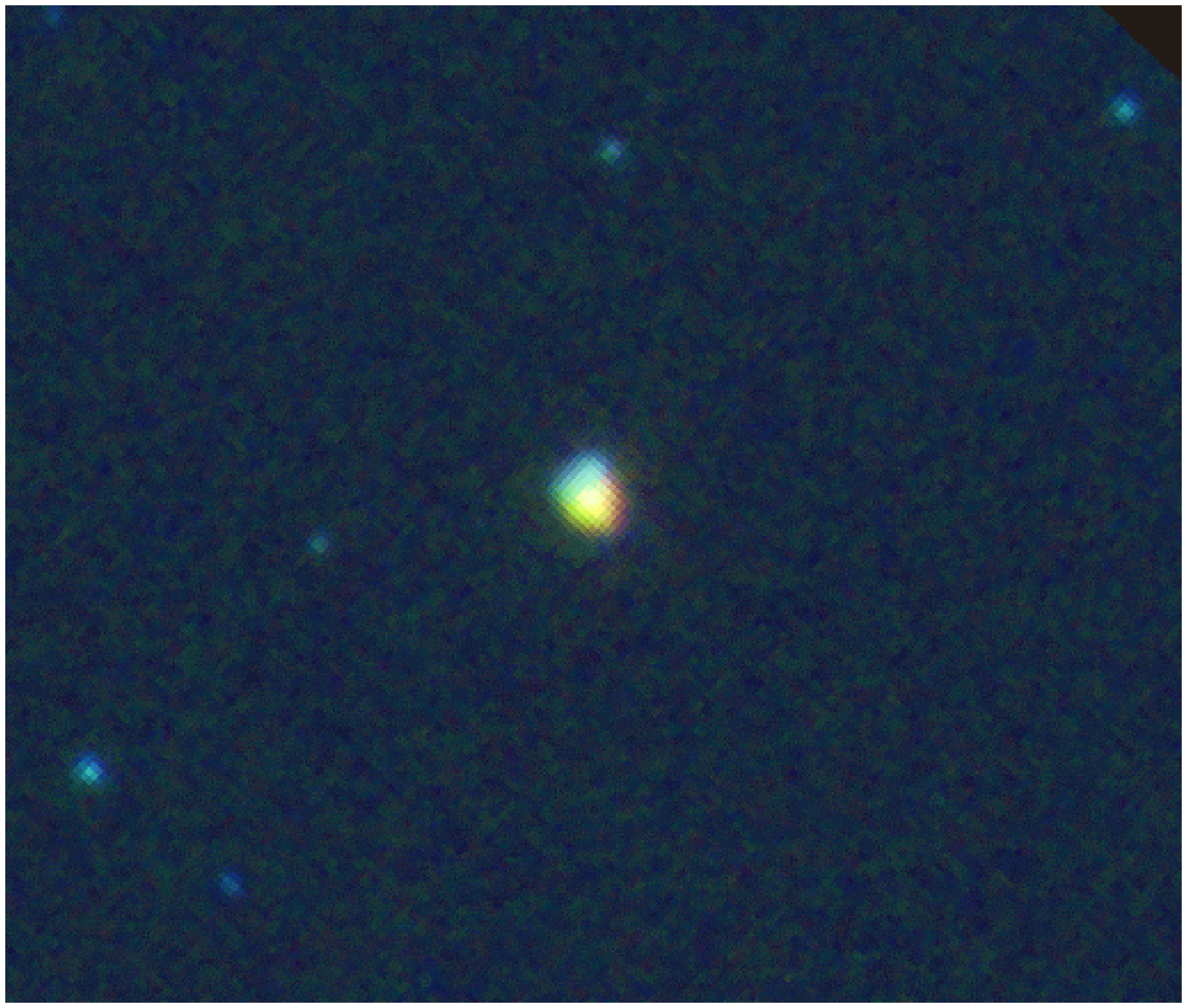}
\figcaption{\footnotesize
{\it Top}: These four \HST\/ images show V4332~Sgr in 1997 (top left: H$\alpha$
filter) and in 2014 (top right: $B$ [F438W]; bottom left: $V$ [F606W]; bottom
right: $I$ [F814W]). Note the nearly stellar image in 1997, about three years
after the outburst, and the resolved ejecta in 2014, 20~years after outburst.
Logarithmic stretches were used, and each frame is $8\farcs3$ high and has north
at the top, east on the left.
{\it Bottom}: Color composite of the 2014 images; red, green, and blue represent
the frames in $I$, $V$, and $B$. This frame is $5\farcs8$ high. The ejecta are
bluer than the central source, consistent with scattering off dust particles.
}
\end{center}
\end{figure}

\end{document}